\def   \ni {\noindent}

\def   \ssk {\vskip  5truept}

\def   \bsk {\vskip 15truept}
 
\def   \newpage {\vfill\eject}
\def   \newline {\hfil\break}

\documentstyle[epsfig]{article}
\begin{document}

\hsize 5truein
\vsize 8truein
\font\abstract=cmr8
\font\keywords=cmr8
\font\caption=cmr8
\font\references=cmr8
\font\text=cmr10
\font\affiliation=cmssi10
\font\author=cmss10
\font\mc=cmss8
\font\title=cmssbx10 scaled\magstep2
\font\alcit=cmti7 scaled\magstephalf
\font\alcin=cmr6 
\font\ita=cmti8
\font\mma=cmr8
\def\ref{\par\noindent\hangindent 15pt}
\null


\title{\ni ASTROPHYSICS WITH HESSI
}                                               

\bsk \bsk
\author{\ni D.M.~Smith $^{1}$, R.P.~Lin $^{1}$, J.~McTiernan $^{1}$, and A.S.~ Slassi-Sennou $^{1}$}                                                       
\bsk
\affiliation{1) Space Sciences Laboratory, University of California, Berkeley, USA
}                                                
\bsk
\baselineskip = 12pt

\abstract{ABSTRACT \ni
In the summer of the year 2000, a NASA Small Explorer
satellite, the High Energy Solar Spectroscopic Imager (HESSI), will be
launched.  It will consist of nine large, coaxial germanium detectors
viewing the Sun through a set of Rotation Modulation Collimators and
will accomplish high-resolution imaging and spectroscopy of solar
flares in the x-ray and gamma-ray bands.  Here we describe some of the
astrophysical observations HESSI will also perform in addition to its
solar mission.
}                                                    
\bsk
\baselineskip = 12pt

\bsk
\baselineskip = 12pt


\text{\ni 1. What is HESSI?
\ssk
\ni     
The High Energy Solar Spectroscopic Imager (HESSI) is a NASA Small
Explorer mission being built at the University of California at
Berkeley (Prof. Robert P. Lin, Principal Investigator), the NASA
Goddard Space Flight Center, the Paul Scherrer Institut in
Switzerland, and Spectrum Astro, Inc., with participation by a number of
other institutions.  It is scheduled for launch into low-Earth
orbit in July 2000.

HESSI's primary science goals are imaging spectroscopy (3 keV to
several MeV), and high-resolution nuclear spectroscopy of solar flares
during the next solar maximum. The instrument consists of 9 large
germanium detectors (cooled to 75 K by a mechanical cooler) which
cover the full energy range.  The detectors sit below a Rotation
Modulation Collimator (RMC) system for high resolution imaging
capability (2 arcsec at hard x-ray energies).  The rotation
is provided by spinning the whole spacecraft at about 15 rpm.

Each 
germanium detector is a closed-end coaxial cylinder with a volume of
over 300 cm$^3$, electronically segmented into a thin front
segment and thick rear segment.  The rear segments view
nearly half the sky through side walls of only 4mm of aluminum, giving
then an effective energy range of 20 keV to 15 MeV.  The front
segments shield the rear segments from solar photons below 100 keV and
view the Sun through beryllium windows and a small amount of
insulating blankets, giving them a useful energy range down to about 3
keV.

\bsk
\ni 2.  HESSI Astrophysics 
\ssk
\ni 
Although HESSI is primarily a solar mission, the HESSI team is
committed to making sure its capabilities for extra-solar
astrophysical observations are fully exploited.  All HESSI data
and analysis software will be public, with no proprietary period.

\newpage

The astrophysics program the HESSI team is planning to pursue combines
aspects of what has been done with the CGRO/BATSE and Wind/TGRS
instruments, as well as techniques unique to HESSI.  In addition to
the topics discussed below, high-resolution spectroscopy of gamma-ray
bursts and soft gamma repeater (SGR) events will be accomplished
without the need for a burst trigger: every photon is always
telemetered with its arrival time.

\bsk
\ni 3. RMC Imaging of the Crab Nebula  
\ssk
\ni 

The Crab nebula and pulsar (ecliptic latitude -1.3$^{\circ}$), will
drift into HESSI's imaged field of view once per year, so we will
automatically produce images from about 3 keV to 100 keV.  Only one
image above a few keV has ever been produced (Pelling et al. 1997),
with a resolution of about 15 arcsec, as compared to HESSI's 2 arcsec.
The ROSAT soft x-ray image of the nebula (Hester et al. 1995) shows
features at this scale, as do the radio wisps, so the hard x-ray image
should be very informative.  Our simulations indicate the statistics
will be good enough to see the relevant details.  In addition, the
radio wisps evolve on the scale of a year, so we may be able to
observe annual changes in the hard x-ray image.

\bsk
\ni 4.  Galactic Gamma-Ray Lines
\ssk
\ni 

By using the Earth as an occulter, we can produce
background-subtracted spectra of the Galactic Center region.  In this
analysis, the whole HESSI array will be treated as a single detector.
Spectra will be summed over a time on the order of a minute (several
revolutions), and background will be constructed from data taken
during other orbits when the Galactic Center was blocked by the Earth.
A similar technique has been used to measure the Galactic 511 keV line
with BATSE to the highest precision of any experiment. (Smith et
al. 1998; see also the other poster by D. M. Smith et al. at this
meeting).

\begin{figure}
\centerline{\psfig{file=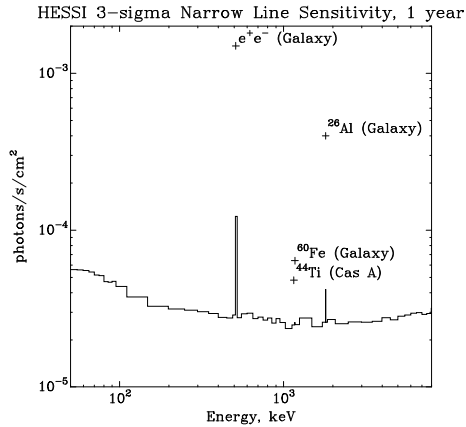, width=7cm}}
\caption{FIGURE 1. HESSI narrow-line sensitivity
}
\end{figure}

Figure 1 shows HESSI's 3$\sigma$ sensitivity to narrow lines in one
year of observations.  The sensitivity is not as good as the INTEGRAL SPI,
since HESSI is unshielded, but HESSI also observes a much larger
portion of the sky at once, and will therefore receive a larger
(albeit unimaged) signal in the diffuse Galactic lines.  This will
make HESSI a good complement to INTEGRAL; subtracting the fluxes
in INTEGRAL line maps from HESSI fluxes will allow us to find
large scale, low-surface-brightness components in the 511 keV and
1809 keV lines.

Important results awaiting confirmation include:
\begin{itemize} \item The small (or zero) amount of Galactic 511 keV flux
which is in the relatively broad, 6.4 keV FWHM component expected from 
annihilation in flight after charge exchange with neutral hydrogen.
This result, from Harris et al. 1998, implies that Galactic positrons are
mostly magnetically excluded from cold cloud cores.
\item The large width (5.4 (+1.4,-1.3) keV FWHM) measured for the 
integrated Galactic 1809 keV line by the GRIS balloon 
instrument (Naya et al. 1996).  This unexpectedly high width means
that $^{26}$Al ejected in supernovae maintains high velocities long
after it would be expected to slow down in the ISM.
\item The low upper limit on $^{60}$Fe, also from GRIS, constraining
models of supernova nucleosynthesis when compared to $^{26}$Al (and
assuming most of the $^{26}$Al is produced in supernovae).
\end{itemize}

\bsk
\ni 5.   Pulsar Period Folding
\ssk
\ni 

\begin{figure}
\centerline{\psfig{file=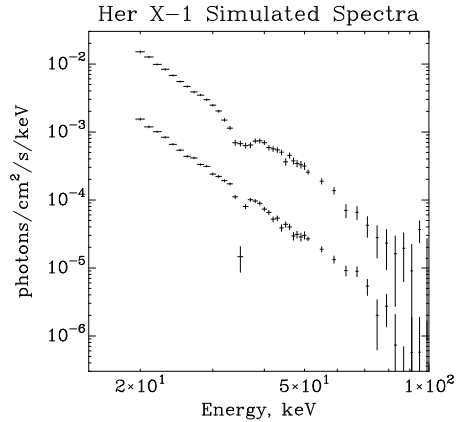, width=7cm}}
\caption{FIGURE 2. Simulated Her X-1 spectra (1 month)
}
\end{figure}

By folding the rear-segment data on the period of known accreting pulsars,
we will produce some of the best high-resolution spectra of the
pulsed emission from these objects.  Figure 2 shows our expected spectrum
from the pulsed emission of Her X-1 in one month of observation.
The upper curve was generated under the assumption that the cyclotron
absorption line is of the same width as the resolution of the scintillators
which have generally observed it.  The lower curve, divided by 10 for
clarity, shows the spectrum we would observe if the absorption line were
narrower than the resolution of HESSI's germanium detectors.

In addition, pulsar period folding will allow us to follow the period
evolution of the sources, a project which has been extremely fruitful
for BATSE (Bildsten et al. 1997).  Although HESSI will have less than
10\% of BATSE's effective area for these observations, there are still
a number of known sources which will be bright enough to follow.
Furthermore, since every photon will be recovered with a time tag,
HESSI will not have the limitation of BATSE's normal operating mode,
which samples rates every 2 seconds.  We will therefore be able to do
a long-term survey of the undersampled range of periods
$<$ 4 sec.

\bsk
\ni 5.   Spin Period Folding
\ssk
\ni 

By folding the rear-segment data on the spin period of the spacecraft,
we will observe bright Galactic point sources by analyzing the
repeated occultation of one detector by another with respect to the
source.  BATSE's success with occultation by the Earth is well known
(Harmon et al. 1992; Zhang et al. 1993).  Although HESSI is much
smaller than BATSE, we have the advantage of gaining many more
occultations per orbit: about 750 detector/detector occultations due
to spin per orbit in addition to the 2 Earth occultations.  We
expect to monitor transients and persistent sources above a few hundred mCrab.

Although the detectors are not completely opaque at 511 keV, we may be
able to obtain some spatial information on the 511 keV line by
detector/detector occultation, in a way similar to the analysis done
by Harris et al. (1998) for Wind/TGRS, but taking advantage of HESSI's
extra order of magnitude of germanium volume.

}

\bsk
\baselineskip = 12pt


{\references \ni REFERENCES
\ssk

\ref Bildsten, L. et al. 1997, ApJ 113, 367
\ref Harmon, B. A. et al. 1992, Proc. Compton Observatory Workshop, p. 69 
\ref Harris, M. J. et al. 1998, ApJ, 501, 55
\ref Hester, J. J. et al. 1995, ApJ, 448, 240
\ref Naya, J. E. et al. 1996, Nature, 384, 44
\ref Pelling, R. M. et al. 1987, ApJ 319, 416

\ref Smith, D. M. et al. 1998, Proc. of the 4th Compton Symposium, AIP Conf.
Proc. \#410, p. 1012
\ref Zhang, S. N. et al. 1993, Nature 366, 245

}                      

\end{document}